\documentclass[a4paper,nobibnotes,nofootinbib,preprintnumbers]{revtex4}
\usepackage{amssymb}
\usepackage{amsmath}
\usepackage{epsfig}
\newcommand{\be}{\begin{equation}}
\newcommand{\ee}{\end{equation}}
\newcommand{\bea}{\begin{eqnarray}}
\newcommand{\eea}{\end{eqnarray}}
\newcommand{\nn}{\nonumber}

\newcommand{\D}{\displaystyle}
\newcommand{\bs}{\boldsymbol}
\newcommand{\g}{\gamma}
\newcommand{\f}{\frac}

\newcommand{\bra}{\langle}
\newcommand{\ket}{\rangle}
\newcommand{\da}{\dagger}

\newcommand\lr[1]{{\left({#1}\right)}}
\begin{document}
\title{Forward inclusive dijet production and azimuthal correlations in pA collisions}
\author{Cyrille Marquet}\email{marquet@quark.phy.bnl.gov}
\affiliation{RIKEN BNL Research Center, Brookhaven National Laboratoty, Upton, NY 11973, USA}
\preprint{RBRC-682}
\begin{abstract}

We derive forward inclusive dijet production in the scattering of a
dilute hadron off an arbitrary dense target, whose partons with small
fraction of momentum $x$ are described by a Color Glass Condensate.
Both multiple scattering and non-linear QCD evolution at small$-x$ are included.
This is of relevance for measurements of two-particle correlations in the
proton direction of proton-nucleus collisions at RHIC and LHC energies.
The azimuthal angle distribution is peaked back to back and broadens as
the momenta of the measured particles gets closer to the saturation scale.

\end{abstract}
\maketitle
\section{Introduction}

Understanding particle correlations in nucleus-nucleus collisions has been
the purpose of many studies in recent years at the Relativistic Heavy Ion
Collider (RHIC). These measurements provide insight on the properties of
dense QCD matter, in particular they could help distinguish to what extend
the experimental observations are due to initial-state or final-state effects. 
For instance, parton saturation \cite{psat} is a characteristic of the initial
nuclear wave functions, and is usually entangled with the final-state phenomenon
of medium-induced energy loss \cite{jetq}. It is so in the suppression of
high-$p_T$ forward hadronic yields. By contrast, with measurements of particle
correlations, initial-state and final-state effects may be disentangled.

As a complement to single inclusive particle spectra, or to nuclear modification
factors, the simplest observables are obtained with two-particle correlations.
In particular, the correlation in azimuthal angle can provide further insight
on the dynamics of energy loss. For instance it has been argued \cite{coneflow}
that the energy radiated off a hard final-state parton is expected to propagate in
such a way that it should lead to an azimuthal correlation featuring a double-peak
structure, instead of the standard back-to-back peak. This has been observed
in Au-Au collisions, the away-side azimuthal correlation splits into a double peak
with maxima displaced away from $180^\circ.$ Meanwhile, the contribution of initial
state saturation effects to the azimuthal decorrelation has yet to be determined.

The main purpose of the present work is to study the inclusive two-particle spectrum
for the following process: $h{\cal T}\!\to\!h_1h_2X$ where $h$ is a dilute hadron and the measured particles $h_1$ and $h_2$ are detected in the very forward direction of that hadron. In this case, only the high-momentum valence partons of $h$ contribute to the scattering and the dominant partonic subprocess is $q{\cal T}\!\to\!qgX:$ a valence quark emits a virtual gluon $g$ via lowest-order pQCD Bremsstrahlung and the quark-gluon fluctuation is put on shell by the interaction with the target ${\cal T}.$ In practice, this is relevant for forward particle production in deuteron-gold ($d\!-\!Au$) collisions at RHIC energies, where only valence quark contribute. For proton-lead ($p\!-\!Pb$) collisions at the LHC, even in the forward region, it is likely that one also needs to account for the $g\!\to\!q\bar q$
and $g\!\to\!gg$ subprocesses.

Meanwhile, it is suited to describe the target ${\cal T}$ by a Color Glass Condensate (CGC), as the process is mainly probing partons with a very small fraction of momentum $x:$ the energetic probes formed by the projectile's valence quark, and quark-gluon fluctuation, interact coherently over the whole longitudinal extension of the target, and see the small$-x$ gluons inside ${\cal T}$ as a dense system of weakly interacting gluons, that behave collectivelly.
It has recently become clear, and it is manifest in the process of interest here, that the CGC cannot be described in terms of a single gluon distribution function, but rather must be described by n-point functions of Wilson-line operators. In other words, the so-called 
$k_T-$factorization used in \cite{klm} is not applicable. We explicitely derive the two-particle spectrum and obtain that it involves up to a 6-point function of Wilson lines in the fundamental representation. In addition, we do not restraint the calculation to the soft-gluon approximation as done in \cite{bknw}.

We obtain a formula similar to that of \cite{nszz}, but their counterpart of our n-point functions are treated differently than in the present paper. In this work, both multiple scattering and non-linear QCD evolution at small$-x$ are included: the Wilson-line
correlators are computed in the framework of the Balitsky-Kovchegov (BK) equation \cite{bk}. 
Considering the correlations in azimuthal angle, we obtain that the perturbative back-to-back peak of the azimuthal angle distribution, which we recover for very large momenta of the measured particles, is reduced by initial state saturation effects. As the momenta decrease closer to the saturation scale $Q_s,$ the angular distribution broadens, but at RHIC and even LHC energies, saturation does not lead to a complete disappearance of the back-to-back peak.

The plan of the paper is as follows. In Section II, the inclusive two-particle spectrum 
$q{\cal T}\!\to\!qgX$ is calculated. The cross-section, differential with respect to the quark and gluon transverse momenta and rapidities, is expressed in terms of Wilson-line correlators in the CGC wavefunction. Section III is devoted to those CGC averages and describes how to perform them in the context of the BK evolution. In Section IV, we introduce the relevant observable to study the azimuthal angle correlation and we investigate the impact of the
small$-x$ evolution at RHIC and LHC energies.

\section{Forward inclusive dijet production cross-section}

In this section we derive the inclusive quark-gluon production cross-section 
in the high-energy scattering of a quark off a Color Glass Condensate. We shall work
at leading order with respect to the strong coupling constant $\alpha_S,$ and to
all orders with respect to $g_S{\cal A}$ where ${\cal A}$ is the CGC color field.
The relevant diagrams are shown in Fig.1 and the necessity of both diagrams will become
transparent in the following derivation: it is a manifestation that there should be
no gluon production without interaction with the target. We will later require that the scattering process features a hard momentum transfer $\Delta\!\gg\!\Lambda_{QCD},$ which
justifies the use of perturbation theory.

We shall use light-cone coordinates with the incoming quark moving along the $x^+$ direction, and work in the light-cone gauge ${\cal A}^+=0.$ In this case, when the quark
passes through the CGC and interacts with its color field, the dominant couplings
are eikonal: the partonic components of the dressed quark wavefunction have frozen
transverse coordinates and the gluonic field of the target does not vary during the interaction. This is justified since the incident projectile propagates at nearly
the speed of light and its time of propagation through the target is shorter
than the natural time scale on which the target fields vary. The effect of the
interaction with the target is that the components of the dressed quark wavefunction
pick up eikonal phases.

\subsection{The dressed quark wavefunction}

To describe the wavefunction of the incoming quark, we shall use light-cone perturbation theory. We denote the $3-$momentum of the quark $p\!=\!(p^+,p_\perp),$ and its spin and color indices $\alpha$ and $i.$ One has $p^-\!=\!(p_\perp^2+m^2)/2p^+.$ To decompose its
wavefunction on bare quark and gluon states, we introduce $b^\da_{i,\alpha}(k)$ and
$b_{i,\alpha}(k)$ (resp. $a^\da_{c,\lambda}(k)$ and $a_{c,\lambda}(k)$), the
creation and annihilation operators of a quark with color $i,$ spin $\alpha$
(resp. gluon with color $c,$ polarization $\lambda$), and $3-$momentum $k.$ One has
\bea
b^\da_{i,\alpha}(k)|0\ket=|k,i,\alpha\ket_0\ ,\hspace{0.5cm}&
b_{i,\alpha}(k)|0\ket=0\ ,\hspace{0.5cm}&
\left\{b_{i,\alpha}(k),b^\da_{j,\beta}(k')\right\}=\delta_{ij}
\delta_{\alpha\beta}\delta^{(3)}(k-k')\ ,\\
a^\da_{c,\lambda}(k)|0\ket=|k,c,\lambda\ket_0\ ,\hspace{0.5cm}&
a_{c,\lambda}(k)|0\ket=0\ ,\hspace{0.5cm}&
\left[a_{c,\lambda}(k),a^\da_{d,\lambda'}(k')\right]=\delta_{cd}
\delta_{\lambda\lambda'}\delta^{(3)}(k-k')\ .
\eea
We recall that in light-cone quantization, the virtuality of the particle comes from the
non-convervation of the momentum in the $x^-$ direction, meaning
$(k\!-\!k')^-\!\neq\!k^-\!-\!k'^-$. All particles are on-shell and only the $3-$momentum is conserved.

We work at lowest order with respect to $\alpha_S$ so we only need to consider the
Fock states $|q\ket_0$ and $|qg\ket_0$ in the decomposition of the dressed quark:
\be
|p,i,\alpha\ket=Z|p,i,\alpha\ket_0+\sum_{j\beta c\lambda}\int d^3k\ g_S
T^c_{ij}\ \psi^{\lambda}_{\alpha\beta}(p,k)\ 
|(p\!-\!k,j,\beta);(k,c,\lambda)\ket_0\ .\label{inmom}\ee
The $|qg\ket_0$ part of the dressed quark is characterized by the wavefunction 
$g_S T^c_{ij}\psi^{\lambda}_{\alpha\beta}$ where $\beta$ and $j$ denote the spin and color indices of the quark and $\lambda$ and $c$ denote the polarization and color indices of the gluon. $T^c$ is the generator of the fundamental representation of $SU(N_c)$ and 
$\psi^{\lambda}_{\alpha\beta}$ is given by
\be
\psi^{\lambda}_{\alpha\beta}(p,k)=\f1{\sqrt{8(p\!-\!k)^+p^+k^+}}
\f{\bar{u}_\beta(p\!-\!k)\gamma_\mu\varepsilon^\mu_{(\lambda)}(k)u_\alpha(p)}
{(p\!-\!k)^- +k^- -p^-}\ee
where $k\!=\!(k^+,k_\perp)$ and $p\!-\!k\!=\!(p^+\!-\!k^+,p_\perp\!-\!k_\perp)$ are the $3-$momenta of the gluon and quark respectively. The factor $Z$ is a renormalization constant determined from the requirement that the normalization of the dressed quark is the same than that of the bare quark (note that those normalizations are proportional to $\delta^{(3)}(0)$ because we are using plane waves). Physically, $Z$ accounts for the virtual gluon emission associated with the real gluon emission explicit in \eqref{inmom}. When calculating inelastic processes, as is the case here, the actual value of $Z$ is not relevant. 

With our choice of gauge, one can write the two gluon polarization vectors 
$\varepsilon^\mu_{(1)}(k)$ and $\varepsilon^\mu_{(2)}(k)$ in terms of two transverse vectors 
$\varepsilon_\perp^1$ and $\varepsilon_\perp^2:$ $\varepsilon^\mu_{(\lambda)}(k)
=(0,\varepsilon_\perp^\lambda,k_\perp\cdot\varepsilon_\perp^\lambda/k^+).$ Then using the chiral representation of the Dirac matrices to obtains the spinors leads to
\be
\psi^{\lambda}_{\alpha\beta}(p,k)=\f1{\sqrt{k^+}}
\f1{(k_\perp\!-\!zp_\perp)^2\!+\!m^2 z^2}\left\{\begin{array}{lll}\D
\sqrt{2}(k_\perp\!-\!zp_\perp)\cdot\varepsilon_\perp^1[\delta_{\alpha-}\delta_{\beta-}
\!+\!(1\!-\!z)\delta_{\alpha+}\delta_{\beta+}]
\!+\!m z^2\delta_{\alpha+}\delta_{\beta-}\hspace{0.5cm}\lambda=1
\\\\\D
\sqrt{2}(k_\perp\!-\!zp_\perp)\cdot\varepsilon_\perp^2[\delta_{\alpha+}\delta_{\beta+}
\!+\!(1\!-\!z)\delta_{\alpha-}\delta_{\beta-}]
\!-\!m z^2\delta_{\alpha-}\delta_{\beta+}\hspace{0.5cm}\lambda=2
\end{array}\right.\label{qtoqgmom}
\ee
with $z\!=\!k^+/p^+.$ For reasons which will become clear later on, it is more convenient to work in a mixed space, in which the transverse momenta are Fourier transformed into 
transverse coordinates:
\be
b^\da_{i,\alpha}(p^+,\textbf{b})=\int d^2b\ e^{-ip_\perp.\textbf{b}}\ 
b^\da_{i,\alpha}(p^+,p_\perp)\ ,\hspace{0.5cm}
a^\da_{c,\lambda}(p^+,\textbf{x})=\int d^2b\ e^{-ip_\perp.\textbf{x}}\ 
a^\da_{c,\lambda}(p^+,p_\perp)\ .\label{mixrep}\ee
In the mixed representation, the decomposition of the dressed quark \eqref{inmom}
on the Fock states $|q\ket_0$ and $|qg\ket_0$ is the following:
\be
|p,i,\alpha\ket=\int\f{d^2b}{(2\pi)^2}\ e^{ip_\perp.\textbf{b}} 
\left[Z|p^+,\textbf{b},i,\alpha\ket_0+\sum_{j\beta c\lambda}\int dk^+ \f{d^2x}{(2\pi)^2}\
g_S T^c_{ij}\ \phi^\lambda_{\alpha\beta}(p,k^+,\textbf{x}\!-\!\textbf{b})\
|(p^+\!-\!k^+,\textbf{b},j,\beta);(k^+,\textbf{x},c,\lambda)\ket_0\right]\label{incoor} \ee
where $\textbf{b}$ and $\textbf{x}$ are the transverse positions of the quark and gluon respectively, and with the mixed-space wavefunction
\be
\phi^\lambda_{\alpha\beta}(p,k^+,\textbf{x})=\int d^2k_\perp\ 
e^{ik_\perp.\textbf{x}}\ \psi^{\lambda}_{\alpha\beta}(p,k)\ee
given by
\be
\phi^\lambda_{\alpha\beta}(p,k^+,\textbf{x})=
\f{2\pi m}{\sqrt{k^+}}\ e^{izp_\perp.\textbf{x}}\left\{\begin{array}{ll}\D
iz\sqrt{2}K_1(mz|\textbf{x}|)\f{\textbf{x}\cdot\varepsilon_\perp^1}{|\textbf{x}|}
[\delta_{\alpha-}\delta_{\beta-}\!+\!(1\!-\!z)\delta_{\alpha+}\delta_{\beta+}]
\!+\!z^2K_0(mz|\textbf{x}|)\delta_{\alpha+}\delta_{\beta-}\hspace{0.5cm}\lambda=1
\\\D
iz\sqrt{2}K_1(mz|\textbf{x}|)\f{\textbf{x}\cdot\varepsilon_\perp^2}{|\textbf{x}|}
[\delta_{\alpha+}\delta_{\beta+}\!+\!(1\!-\!z)\delta_{\alpha-}\delta_{\beta-}]
\!-\!z^2K_0(mz|\textbf{x}|)\delta_{\alpha-}\delta_{\beta+}\hspace{0.5cm}\lambda=2
\end{array}\right.\ .\label{qtoqgmix}
\ee

\begin{figure}[t]
\begin{center}
\epsfig{file=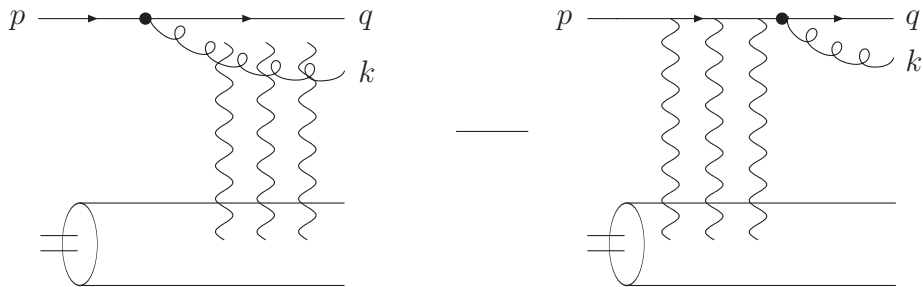,width=12cm}
\caption{Inclusive quark-gluon production cross-section in the high-energy scattering of a quark off a Color Glass Condensate. $p\!=\!(p^+,p_\perp):$ momentum of the incoming quark;
$q\!=\!(q^+,q_\perp)$ and $k\!=\!(k^+,k_\perp):$ momentum of the outgoing quark and gluon.
The vertical wavy lines represent the interaction with the target CGC, each line carries a factor $g_S{\cal A}$ and multiple gluon exchanges must be resummed. The black points represent the emission of the produced gluon by the quark, it is emitted before the interaction or after the interaction in which case the contribution comes with a minus sign, as explained in the text.}
\end{center}
\label{F1}
\end{figure}

\subsection{High-energy eikonal scattering off the target}

Let us first recall the basics of the CGC description. When probing inside a target
hadron with processes that are sentitive to partons with a very small fraction of
momentum $x,$ the probe actually sees a dense system of gluons, responsible for
large classical color fields ${\cal A}\!\sim\!1/g_S$ \cite{mv}. Rather than using a
Fock-state decomposition which is not adapted to account for the collective behavior 
of the small$-x$ gluons, it is more appropriate to use other degrees of freedom and
describe the target by classical color fields:
\be
|{\cal T}\ket=|qqq\ket_0+|qqqg\ket_0+\dots+|qqqg\dots ggg\ket_0+\dots\hspace{0.5cm}
\Rightarrow\hspace{0.5cm}|{\cal T}\ket=\int D{\cal A}\ 
\Phi_{x_A}[{\cal A}]\ |{\cal A}\ket
\label{cgc}\ .\ee
The CGC wavefunction $\Phi_{x_A}[{\cal A}]$ is normalized such that 
$\int D{\cal A}\left|\Phi_{x_A}[{\cal A}]\right|^2=1,$ and $x_A$ denotes 
the smallest fraction of longitudinal momentum probed. It depends on the final-state
kinematics of the process condidered and will be specified later.
The CGC wavefunction is mainly a non-perturbative quantity, but the $x_A$ evolution of 
$\left|\Phi_{x_A}[{\cal A}]\right|^2$ can be computed from perturbative QCD \cite{jimwlk},
in the leading$-\ln(1/x_A)$ approximation that resums powers of $\alpha_S\ln(1/x_A).$
It is a priori not obvious that this description of the target, which requires small values
of $x_A,$ is valid for experiments at present energies. However, it has had success for many observables in the context of HERA \cite{myrev} and RHIC \cite{jyrev}.

The target is moving along the light-cone in the $x^-$ direction, and the only component
of its color current is $J^-.$ With our choice of gauge ${\cal A}^+\!=\!0,$ the current conservation law $[D_\mu,J^\mu]\!=\!\partial^+J^-\!=\!0$ implies that $J^-$ does not depend on $x^-.$ Therefore one writes
\be
J^\mu(x^\nu)=\delta^{\mu -}J^-(x^+,\textbf{x})=\delta^{\mu -}T^c\rho_c(x^+,\textbf{x})\ee
where we have introduced the color charge density of the target $\rho_c.$ Solving the Yang-Mills equations $[D_\mu,F^{\mu\nu}]\!=\!J^\nu$ then leads to (see for instance \cite{yacine}) 
\be
A^\mu(x^\nu)=\delta^{\mu -}T^c{\cal A}_c^-(x^+,\textbf{x})\ ,\hspace{1cm}
-\bs{\nabla}^2{\cal A}_c^-(x^+,\textbf{x})=\rho_c(x^+,\textbf{x})\ .\label{ymsol}\ee
The formal functional integration in \eqref{cgc} stands for the $A^-$ integration.

In a scattering process, the outgoing state is obtained from the incoming state by action of the scattering matrix ${\cal S}.$ When high-energy partons scatter off the CGC, the interaction is eikonal and the ${\cal S}$ matrix acts on quarks and gluons as (see for example
\cite{kovwie}):
\be
{\cal S}|\textbf{b},i\ket\!\otimes\!|{\cal A}\ket=\sum_j W^{ij}_F[{\cal A}](\textbf{b})
|\textbf{b},j\ket\!\otimes\!|{\cal A}\ket\ ,\hspace{1.5cm}
{\cal S}|\textbf{x},c\ket\!\otimes\!|{\cal A}\ket=\sum_{d}W_A^{cd}[{\cal A}](\textbf{x})
|\textbf{x},d\ket\!\otimes\!|{\cal A}\ket\ ,\label{eik}\ee
where the phase shifts due to the interaction are described by $W_F$ and $W_A,$ 
the eikonal Wilson lines in the fundamental and adjoint representations 
respectively, corresponding to propagating quarks and gluons. They are given by
\be
W_{F,A}[{\cal A}](\textbf{x})={\cal P}\exp
\lr{ig_S\int dx^+T_{F,A}^c{\cal A}^-_c(x^+,\textbf{x})}
\ee
with $T_{F,A}^a$ the generators of $SU(N_c)$ in the fundamental ($F$) or adjoint ($A$)
representations and with ${\cal P}$ denoting an ordering in $x^+.$ The Wilson lines resum
powers of $g_S{\cal A}$ which is necessary as ${\cal A}$ is a large classical field whose strength is of order $1/g_S.$ It is manifest from \eqref{eik} why the mixed representation introduced earlier is convenient: working with transverse space coordinates (instead of transverse momenta) provides eigenstates of the high-energy ${\cal S}-$matrix.

Coming back to our computation, the incoming and outgoing states are
\be
|\Psi_{in}\ket=|p,i,\alpha\ket\otimes|{\cal T}\ket\ ,\hspace{1cm}
|\Psi_{out}\ket={\cal S}|\Psi_{in}\ket\ .
\ee
Using \eqref{incoor} and \eqref{eik}, it is straightforward to obtain (we now keep the 
${\cal A}$ dependence of the Wilson lines $W_{F,A}$ implicit)
\bea
|\Psi_{out}\ket=\int D{\cal A}\ \Phi_{x_A}[{\cal A}]\int\f{d^2b}{(2\pi)^2}\
e^{ip_\perp.\textbf{b}} 
\left(Z\sum_j[W_F(\textbf{b})]_{ij}|p^+,\textbf{b},j,\alpha\ket_0\otimes|{\cal A}\ket+
\hspace{5cm}\right.\nonumber\\\left.\sum_{j\beta d\lambda}\int dk^+ \f{d^2x}{(2\pi)^2}\
g_S\phi^\lambda_{\alpha\beta}(p,k^+,\textbf{x}\!-\!\textbf{b})
[T^cW_F(\textbf{b})]_{ij}W_A^{cd}(\textbf{x})\
|(p^+\!-\!k^+,\textbf{b},j,\beta);(k^+,\textbf{x},d,\lambda)\ket_0\otimes|{\cal A}\ket\right)\ .\label{outint}\eea

In this formula, the quark-gluon contribution represents the first contribution 
pictured in Fig.1, for which the gluon is emitted before the interaction. The second contribution, for which the gluon is emitted after the interaction, is hidden in the quark contribution $Z|p^+,\textbf{b},i,\alpha\ket_0.$ To see that, let us write this in
terms of $|p,i,\alpha\ket:$ from formula \eqref{incoor}, and using the fact that the $p_\perp$ dependence of $\phi^\lambda_{\alpha\beta}(p,k^+,\textbf{x})$ is 
$\exp{(ik^+ p_\perp.\textbf{x}/p^+)},$ one obtains:
\bea
Z|p^+,\textbf{b},i,\alpha\ket_0=\int d^2p_\perp\ e^{-ip_\perp.\textbf{b}}|p,i,\alpha\ket
-\sum_{j\beta c\lambda}\int dk^+ \f{d^2x}{(2\pi)^2}d^2b'\ 
\delta(\textbf{b}\!-\!\textbf{b}'\!-\!k^+(\textbf{x}\!-\!\textbf{b}')/p^+) 
g_S T^c_{ij}\ e^{ip_\perp.(\textbf{b}'-\textbf{b})}\nonumber\\
\phi^\lambda_{\alpha\beta}(p,k^+,\textbf{x}\!-\!\textbf{b}')\
|(p^+\!-\!k^+,\textbf{b}',j,\beta);(k^+,\textbf{x},c,\lambda)\ket_0\ .
\label{inint}\eea
One sees that the emission-after-interaction term arises with a 
minus sign. The dressed quark contribution $|p,i,\alpha\ket$ does not contribute to gluon production, and can be removed from the final answer. Indeed, when computing gluon production, gluons which dress the final-state quark should not be included, as these are not actually produced. 

Combining \eqref{outint} and \eqref{inint}, the outgoing state can be simply rewritten as
\be
|\Psi_{out}\ket=\int D{\cal A}\ \Phi_{x_A}[{\cal A}]
\sum_{j\beta c\lambda}\int dk^+ \f{d^2x}{(2\pi)^2}
\f{d^2b}{(2\pi)^2}\ e^{ip_\perp.\textbf{b}} 
g_S\ \Phi^{c\lambda}_{\alpha\beta,ij}(p,k^+,\textbf{x},\textbf{b})\
|(p^+\!-\!k^+,\textbf{b},j,\beta);(k^+,\textbf{x},c,\lambda)\ket\otimes|{\cal A}\ket
\label{outfin}\ee
with
\be
\Phi^{c\lambda}_{\alpha\beta,ij}(p,k^+,\textbf{x},\textbf{b})=
\phi^\lambda_{\alpha\beta}(p,k^+,\textbf{x}\!-\!\textbf{b})
\left[T^dW_F(\textbf{b})W_A^{dc}(\textbf{x})
-W_F(\textbf{b}\!+\!k^+(\textbf{x}\!-\!\textbf{b})/p^+)T^c\right]_{ij}\ .\ee
Once again, the different contributions contained in this wavefunction have a 
straightforward physical meaning: the term containing $W_A$ corresponds to the interation happening after the gluon emission while the contribution without $W_A$ corresponds to the interation taking place before.

\subsection{The inclusive quark-gluon production cross-section 
$\sigma^{q{\cal T}\to qgX}$}

From the outgoing state \eqref{outfin}, one can now compute the production
of a quark with momentum $q=(q^+,q_\perp)$ and a gluon with momentum $k=(k^+,k_\perp).$
The corresponding cross-section cross-section reads
\be
\f{d\sigma^{q{\cal T}\to qgX}}{d^3kd^3q}=\f1{2N_c}\sum_{i\alpha}
\bra\Psi_{out}|N_q(q)N_g(k)|\Psi_{out}\ket
\label{csec}\ee
where we recall that $i$ and $\alpha$ refer to the color and spin of the incoming quark
of momentum $p=(p^+,p_\perp).$ In \eqref{csec}, the operators $N_q$ and $N_g$ are given by
\be
N_q(q)=\sum_{j\beta}b^\da_{j,\beta}(q)b_{j,\beta}(q)\ ,\hspace{1cm}
N_g(k)=\sum_{a\lambda}a^\da_{a,\lambda}(k)a_{a,\lambda}(k)
\ee
in terms of the creation and annihilation operators introduced earlier in Section II-A.
Let us rewrite the cross-section \eqref{csec} using operators in the mixed representation 
\eqref{mixrep}:
\bea
\f{d\sigma^{q{\cal T}\to qgX}}{d^3kd^3q}=\f1{2N_c}\int
\f{d^2x}{(2\pi)^2}\f{d^2x'}{(2\pi)^2}\f{d^2b}{(2\pi)^2}\f{d^2b'}{(2\pi)^2}\ 
e^{ik_\perp.(\textbf{x}'-\textbf{x})}e^{iq_\perp.(\textbf{b}'-\textbf{b})}\nonumber\\
\sum_{a\lambda\alpha\beta ij}\bra\Psi_{out}|
a^\dagger_{a,\lambda}(\textbf{x}',k^+)b^\dagger_{j,\beta}(\textbf{b}',q^+)
a_{a,\lambda}(\textbf{x},k^+)b_{j,\beta}(\textbf{b},q^+)|\Psi_{out}\ket\ ,
\label{csmix}\eea
where $\textbf{x}$ and $\textbf{x}'$ (resp. $\textbf{b}$ and $\textbf{b}'$) represent now the transverse coordinates of the measured gluon (resp. quark) in the amplitude and the complex conjugate amplitude respectively. When computing the action of $a_{a,\lambda}$ and 
$b_{j,\beta}$ on $|\Psi_{out}\ket,$ one obtains
\be
a_{a,\lambda}(\textbf{x},k^+)b_{j,\beta}(\textbf{b},q^+)|\Psi_{out}\ket=g_S\
e^{ip_\perp.\textbf{b}}\ \delta(p^+\!-\!k^+\!-\!q^+)\int D{\cal A}\ \Phi_{x_A}[{\cal A}]\ 
\Phi^{a\lambda}_{\alpha\beta,ij}(p,k^+,\textbf{x},\textbf{b})|{\cal A}\ket\label{abpsiout}\ .\ee
The function $\delta(p^+\!-\!k^+\!-\!q^+)$ in \eqref{abpsiout} shows that the longitudinal momenta are converved during in the high-energy eikonal scattering. However, this delta function at the level of the amplitude will lead to a factor $\delta(0)$ when computing the cross-section \eqref{csmix}. This problem is related to the factor $\delta^{(3)}(0)$ present in the normalization of the state $|\Psi_{in}\ket,$ itself due to the fact that we are using plane waves. When computing a physical observable, this divergence usually goes away, as is the case for the two transverse dimensions. However, a $\delta(0)$ remains for the longitudinal direction precisely because longitudinal momenta are conserved by the interaction. 
Working with wave packets would solve the problem, and the appropriate prescription is to
remove the factor $2\pi\delta(p^+\!-\!k^+\!-\!q^+)$ from the amplitude \eqref{abpsiout}, and to put it back in the cross-section \eqref{csmix}.

Denoting $z\!=\!k^+/p^+$ and using \eqref{abpsiout} in \eqref{csmix}, one obtains
the $q{\cal T}\!\to\!qgX$ cross-section ($\alpha_S\!=\!g_S^2/4\pi$):
\bea
\f{d\sigma^{q{\cal T}\to qgX}}{d^3kd^3q}=\alpha_S C_F
\delta(p^+\!-\!k^+\!-\!q^+)
\int\f{d^2x}{(2\pi)^2}\f{d^2x'}{(2\pi)^2}\f{d^2b}{(2\pi)^2}\f{d^2b'}{(2\pi)^2}\ 
e^{ik_\perp.(\textbf{x}'-\textbf{x})}e^{i(q_\perp-p_\perp).(\textbf{b}'-\textbf{b})}\nonumber\\
\sum_{\lambda\alpha\beta}
\phi^{\lambda^*}_{\alpha\beta}(p,k^+,\textbf{x}'\!-\!\textbf{b}')
\phi^{\lambda}_{\alpha\beta}(p,k^+,\textbf{x}\!-\!\textbf{b})
\left\{S_{qg\bar q g}^{(4)}[\textbf{b},\textbf{x},\textbf{b}',\textbf{x}';x_A]
-S_{qg\bar q}^{(3)}[\textbf{b},\textbf{x},\textbf{b}'\!+\!z(\textbf{x}'\!-\!\textbf{b}');x_A]
\right.\nonumber\\\left.
-S_{qg\bar q}^{(3)}[\textbf{b}\!+\!z(\textbf{x}\!-\!\textbf{b}),\textbf{x}',\textbf{b}';x_A]
+S_{q\bar q}^{(2)}[\textbf{b}\!+\!z(\textbf{x}\!-\!\textbf{b})
,\textbf{b}'\!+\!z(\textbf{x}'\!-\!\textbf{b}');x_A]\right\}\ .
\label{qTtoqgX}\eea
In \eqref{qTtoqgX}, we have introduced the following traces of products of Wilson lines:
\be
S_{qg\bar q g}^{(4)}(\textbf{b},\textbf{x},\textbf{b}',\textbf{x}';x_A)=
\f1{C_FN_c}\left\bra\mbox{Tr}\lr{W_F(\textbf{b})W^\dagger_F(\textbf{b}')T^dT^c}
[W_A(\textbf{x})W_A^\dagger(\textbf{x}')]^{cd}\right\ket_{x_A}\ ,\label{Sqgqg}\ee
\be
S_{qg\bar q}^{(3)}(\textbf{b},\textbf{x},\textbf{b}';x_A)
=\f1{C_FN_c}\left\bra\mbox{Tr}\lr{W^\dagger_F(\textbf{b}')T^cW_F(\textbf{b})T^d}
W_A^{cd}(\textbf{x})\right\ket_{x_A}\ ,\label{Sqgq}\ee
\be
S_{q\bar q}^{(2)}(\textbf{b},\textbf{b}';x_A)=\f1{N_c}\left\bra\mbox{Tr}
\lr{W_F(\textbf{b})W^\dagger_F(\textbf{b}')}\right\ket_{x_A}\ ,\label{Sqq}\ee
and we have also denoted the average over the CGC wavefunction squared
$|\Phi_{x_A}[{\cal A}]|^2$ using the following notation:
\be
\int D{\cal A}\ |\Phi_{x_A}[{\cal A}]|^2 f[{\cal A}]=\left\bra f\right\ket_{x_A}\ .
\label{avg}\ee
The quantities $S_{q\bar q}^{(2)},$ $S_{qg\bar q}^{(3)}$ and $S_{qg\bar q g}^{(4)}$ contain the QCD evolution toward small values of $x_A.$

Some comments about formula \eqref{qTtoqgX} are in order. 
\begin{itemize}
\item First one recovers the result of \cite{nszz}, except that in our case the interaction with the target is obtained by specific $n-$point functions, expressed in terms of Wilson lines that resum powers of $g_S{\cal A}.$ These are to be averaged with the CGC wavefunction squared
$|\Phi_{x_A}[{\cal A}]|^2,$ whose $x_A$ evolution resums powers of $\alpha_S\ln(1/x_A),$ in
the leading$-\ln(1/x_A)$ approximation.
\item Using the Fierz identities
\bea
\left[W_F(\textbf{x})\right]_{ij}[W_F^{\dagger}(\textbf{x})]_{kl}
&=&\frac{1}{N_c}\ \delta_{il}\delta_{jk}
+2W_A^{cd}(\textbf{x})T^c_{il}T^d_{kj}\ ,\label{fierz1}
\\T^c_{ij}T^c_{kl}&=&\f12\delta_{il}\delta_{jk}
-\f1{2N_c}\delta_{ij}\delta_{kl}\ ,\label{fierz2}\eea
one obtains that $W_A^{cd}(\textbf{x})=2\mbox{Tr}(W_F^\da(\textbf{x})T^cW_F(\textbf{x})T^d)$
which shows that an adjoint Wilson line $W_A$ is equivalent to two fundamental Wilson lines $W_F.$ Therefore the quantities $S_{q\bar q}^{(2)},$ $S_{qg\bar q}^{(3)}$ and
$S_{qg\bar q g}^{(4)}$ are 2-, 4- and 6-point functions with respect to the averaging 
\eqref{avg}. Using \eqref{fierz1} and \eqref{fierz2}, one can also see that the singularities
of $\phi^{\lambda}_{\alpha\beta}$ in \eqref{qTtoqgX} are cancelled when
$\textbf{x}\!=\!\textbf{b}$ or $\textbf{x}'\!=\!\textbf{b}'.$
\item Our result should be identical to that of \cite{jamalyuri}, although this is not so straightforward to see, as their expression for the $q{\cal T}\to qgX$ cross-section is not as compact as our formula \eqref{qTtoqgX}. This is probably because in \cite{jamalyuri}, the Wilson lines are expressed in momentum space, where the interaction is not diagonal.
\end{itemize}

Let us finally consider the soft-gluon approximation $z\!\ll\!1.$ Not only does the 
wavefunction \eqref{qtoqgmix} simplify, but also many simplifications occur with the Wilson lines. Indeed using the Fierz identity to transform all the adjoint Wilson lines into fundamental ones, one obtains (if one also factorizes the remaining traces, meaning
$\bra\mbox{Tr}(\ .\ )\mbox{Tr}(\ .\ )\ket_{x_A}=
\bra\mbox{Tr}(\ .\ )\ket_{x_A}\bra\mbox{Tr}(\ .\ )\ket_{x_A},$ we recover the expression of \cite{bknw}):
\bea
S_{qg\bar q g}^{(4)}(\textbf{b},\textbf{x},\textbf{b}',\textbf{x}';x_A)
-S_{qg\bar q}^{(3)}(\textbf{b},\textbf{x},\textbf{b}';x_A)
-S_{qg\bar q}^{(3)}(\textbf{b},\textbf{x}',\textbf{b}';x_A)
+S_{q\bar q}^{(2)}(\textbf{b},\textbf{b}';x_A)=\f{N_c}{2C_F}\left\bra
\f1{N_c}\mbox{Tr}\lr{W_F(\textbf{b})W^\dagger_F(\textbf{b}')}\right.\nn\\
-\f1{N_c}\mbox{Tr}\lr{W_F(\textbf{b})W^\dagger_F(\textbf{x})}
\f1{N_c}\mbox{Tr}\lr{W_F(\textbf{x})W^\dagger_F(\textbf{b}')}
-\f1{N_c}\mbox{Tr}\lr{W_F(\textbf{b})W^\dagger_F(\textbf{x}')}
\f1{N_c}\mbox{Tr}\lr{W_F(\textbf{x}')W^\dagger_F(\textbf{b}')}\nn\\\left.
+\f1{N_c}\mbox{Tr}\lr{W_F(\textbf{x})W^\dagger_F(\textbf{x}')}
\f1{N_c}\mbox{Tr}\lr{W_F(\textbf{b})W^\dagger_F(\textbf{b}')
W_F(\textbf{x}')W^\dagger_F(\textbf{x})}
\right\ket\ .
\eea
The last term in this expression shows that even in the soft-gluon approximation, the dipole degrees of freedom (i.e. traces of two Wilson lines) are not sufficient to compute the dijet cross-section \eqref{qTtoqgX}. This is an important difference with the case of single-particle production \cite{gprod}, for which using dipoles is enough, yielding the so-called
$k_T-$factorization. For two-particle production, $k_T-$factorization cannot be used.

In the following we shall not consider the soft-gluon approximation, and work directly with the 
2-, 4- and 6-point functions \eqref{Sqq}, \eqref{Sqgq} and \eqref{Sqgqg}. Indeed, we are interested in final-state configurations where the two particles are both detected in the forward hemisphere, with similar rapiditites. By contrast, using a soft gluon would impose the constraint of a large rapidity between the two detected hadrons.

\section{Performing the target averages}

In this section, we explain how to model the CGC wavefunction in order to perform the average
\eqref{avg} while taking into account the small$-x$ QCD evolution. To compute the correlators $S_{q\bar q}^{(2)},$ $S_{qg\bar q}^{(3)}$ and $S_{qg\bar q g}^{(4)},$ we shall model the CGC wavefunction squared $\left|\Phi_{x_A}[{\cal A}]\right|^2$ using Gaussian-distributed sources. Then, following the approach of \cite{fgv}, we will work in the large$-N_c$ limit which allows to easily implement the Balitsky-Kovchegov (BK) evolution, and also simplifies the analytic expression for the 6-point function.

\subsection{A Gaussian distribution of sources}

To compute the average \eqref{avg}, we use the following Gaussian distribution of sources
\be
|\Phi_{x_A}[{\cal A}]|^2=\exp\lr{-\int d^2xd^2y dz^+
\f{\rho_c(z^+,\textbf{x})\rho_c(z^+,\textbf{y})}{2\mu_{x_A}^2(z^+,\textbf{x}-\textbf{y})}}
\label{gauss}\ee
where the color charge density $\rho_c$ and the color field ${\cal A}$ are simply related via formula \eqref{ymsol}. The variance $\mu^2_{x_A}$ is a function of $x_A$ and caracterizes the density of the color charges. With this Gaussian approximation, the 2-point and 4-point correlators \eqref{Sqq} and \eqref{Sqgq} have been computed for arbitrary $N_c$ (see for instance the Appendix A of \cite{bgv}). They are given by
\bea
S_{q\bar q}^{(2)}(\textbf{b},\textbf{b}';{x_A})&=&
e^{-\f{C_F}2\Gamma(\textbf{b}-\textbf{b}',x_A)}\ ,\label{2pf}\\
S_{qg\bar q}^{(3)}(\textbf{b},\textbf{x},\textbf{b}';x_A)&=&
e^{-\f{N_c}4[\Gamma(\textbf{x}-\textbf{b},x_A)+\Gamma(\textbf{x}-\textbf{b}',x_A)]
+\f1{4N_c}\Gamma(\textbf{b}-\textbf{b}',x_A)}\ ,\label{4pf}
\eea
with the function $\Gamma(\textbf{b}\!-\!\textbf{b}',x_A)$ related to $\mu^2_{x_A}$ in the following way
\be
\Gamma(\textbf{b}-\textbf{b}',x_A)=g_S^4\int d^2xd^2y dz^+\
\mu_{x_A}^2(z^+,\textbf{x}-\textbf{y})
[G_0(\textbf{b}-\textbf{x})-G_0(\textbf{b}'-\textbf{x})]
[G_0(\textbf{y}-\textbf{b})-G_0(\textbf{y}-\textbf{b}')]\ ,\label{bksol}
\ee
where $G_0$ is the two-dimensional massless propagator
\be
G_0(\textbf{x})=\int\limits_{|\textbf{k}|>\Lambda_{QCD}}\!\f{d^2k}{{(2\pi)^2}}
\f{e^{i\textbf{k}\cdot\textbf{x}}}{\textbf{k}^2}\ .
\ee

The last correlator needed for our study has a more complicated structure. Following the derivation of the 4-point function in \cite{bgv}, it is easy to see that the problem of computating the six point function \eqref{Sqgqg} can be reduced to the diagonalisation of a
$6\times6$ matrix. Note that in the recent publication \cite{kenyos}, a method to compute any $n-$point function is presented. As already mentioned, only the large$-N_c$ limit is used in the implementation of the small$-x$ QCD evolution. Therefore, computing the exact structure of the 6-point function
$S_{qg\bar q g}^{(4)}$ is of no interest for the present study. We only give the large$-N_c$ result
\be
S_{qg\bar q g}^{(4)}(\textbf{b},\textbf{x},\textbf{b}',\textbf{x}';x_A)=
e^{-\f{N_c}4[\Gamma(\textbf{x}-\textbf{b},x_A)+\Gamma(\textbf{x}'-\textbf{b}',x_A)
+\Gamma(\textbf{x}-\textbf{x}',x_A)]}\ .\label{6pf}
\ee
This will be used along with formula \eqref{2pf} and the large$-N_c$ limit of formula
\eqref{4pf}. Note that these expression should be understood for scattering at fixed impact parameter, which is why they feature one less independent variable than excepted. We will assume in the following that the impact-parameter dependence of the correlators factorizes.

\subsection{Evolving the MV model with the BK equation}

Let us now explain the strategy to evaluate the function $\Gamma(\textbf{r},x_A),$ in terms of which all the correlators \eqref{2pf}, \eqref{4pf} and \eqref{6pf} could be written. We will use the fact that, under the asumption \eqref{gauss} that the CGC wavefunction is Gaussian, the large$-N_c$ limit implies the following result
\be
\left\bra\mbox{Tr}
\lr{W_F(\textbf{b})W^\dagger_F(\textbf{b}')}
\mbox{Tr}
\lr{W_F(\textbf{x})W^\dagger_F(\textbf{x}')}\right\ket_{x_A}=
\left\bra\mbox{Tr}
\lr{W_F(\textbf{b})W^\dagger_F(\textbf{b}')}\right\ket_{x_A}
\left\bra\mbox{Tr}
\lr{W_F(\textbf{x})W^\dagger_F(\textbf{x}')}\right\ket_{x_A}\ .\label{fact}
\ee
This significantly simplifies the high-energy QCD evolution equations. Indeed, when considering the Balitsky hierarchy of equations for the $n-$point correlators, (which is a rewritting the
JIMWLK functional equation for the evolution of $|\Phi_{x_A}[{\cal A}]|^2$ with $x_A$), the factorization \eqref{fact} reduces the QCD evolution to a single closed non-linear equation for $S_{q\bar q}^{(2)}(\textbf{b},\textbf{b}';x_A),$ known as the BK equation \cite{bk}. In this work, we consider the impact-parameter independent version
\be
\f{d S_{q\bar q}^{(2)}(\textbf{b}-\textbf{b}';x)}{d\ln\lr{1/x}}=\bar\alpha\int\f{d^2z}{2\pi}
\f{(\textbf{b}-\textbf{b}')^2}{(\textbf{b}-\textbf{z})^2(\textbf{z}-\textbf{b}')^2}
\lr{S_{q\bar q}^{(2)}(\textbf{b}-\textbf{z};x)S_{q\bar q}^{(2)}(\textbf{z}-\textbf{b}';x)
-S_{q\bar q}^{(2)}(\textbf{b}-\textbf{b}';x)}\label{bk}\ee
with $\bar\alpha\!=\!\alpha_S N_c/\pi.$ From formula \eqref{bksol}, the solution of the BK equation \eqref{bk} gives $\Gamma(\textbf{b}-\textbf{b}',x_A),$ which allows to compute all the $n-$point functions needed for the calculation of the $q{\cal T}\!\to\!qgX$ cross-section.

As the initial condition, we shall use the McLerran-Venugopalan (MV) model \cite{mv} for
$\Gamma(\textbf{r},x_0).$ It is obtained from a Gaussian average of the type \eqref{gauss}, with $\mu_{x_0}^2(z^+,\textbf{x})=\delta(\textbf{x})\mu_{x_0}^2(z^+):$
\be
\Gamma(\textbf{r},x_0)=g_S^4 \textbf{r}^2\lr{\int dz^+\ \mu_{x_0}^2(z^+)}
\int_{|\textbf{r}|\Lambda_{QCD}}^\infty dk\ \f{1-J_0(k)}{\pi k^3}\ .
\label{mvmod}\ee
For $|\textbf{r}|\Lambda_{QCD}\ll 1,$ the remaining integral behaves as
$-\ln(\textbf{r}^2\Lambda^2_{QCD})/(8\pi).$ This leads to the MV model:
\be
\Gamma(\textbf{r},x_0)=\f1{2C_F}\ \textbf{r}^2\ Q_{s_0}^2
\ln\lr{e+\f1{\textbf{r}^2\Lambda^2_{QCD}}}\ ,\hspace{0.5cm} 
Q_{s_0}^2=\f{g_S^4 C_F}{4\pi} \lr{\int dz^+\ \mu_{x_0}^2(z^+)}
\ee
where $Q_{s_0}$ is the initial saturation scale, at $x\!=\!x_0.$ In practice, we choose
to start the BK evolution at $x_0\!=\!0.01,$ which is small enough to justify using the MV model. For a target Gold or Lead nucleus, $2\pi Q^2_{s_0}\!=\!2\ \mbox{GeV}^2$ is appropriate at $x_0=0.01$ (we use the same value as in \cite{fgv}, our saturation scales differ by a
factor $\sqrt{2\pi}$). For that moderately small value of $x,$ the MV model gives reasonable results in heavy ion collisions at RHIC \cite{jyrev}. For $x\!<\!x_0,$ the quantum evolution effects are then implemented by the BK evolution.

It is known that the BK equation \eqref{bk}, equiped with the leading-logarithmic
\cite{bfkl} Balitsky-Fadin-Kuraev-Lipatov (BFKL) kernel
$\chi(\g)\!=\!2\psi(1)\!-\!\psi(\g)\!-\!\psi(1\!-\!\g)$ (in Mellin space), leads to an increase of the saturation scale that goes as
\be
Q_s^2(x)=Q^2_{s_0}\lr{\f{x_0}{x}}^{v\bar\alpha}\mbox{ with }
v=\chi'(\g_c)=\f{\chi(\g_c)}{\g_c}=4.88\hspace{0.5cm}(\g_c=0.6275)\ .
\ee
In practice however, the appropriate saturation exponent is $Q_s^2\!\sim\!x^{-\lambda}$ with 
$\lambda\!\simeq\!0.25,$ and this discrepency is understood in terms of subleading logarithms
\cite{nllsat}. In this work, in order to mimic to correct evolution of the saturation scale with $x,$ we impose the unphysical value $\bar\alpha\!=\!0.05$ when solving the BK equation
\eqref{bk}. This is a simple way to account for the unknown next-leading effects, while staying compatible with experimental observations.

\subsection{Expression for the cross-section}

We now come back to the expression \eqref{qTtoqgX} for the $q{\cal T}\!\to\!qgX$ cross-section, which can be simplified using the results \eqref{2pf}, \eqref{4pf} and \eqref{6pf}
for the 2-, 4- and 6-point functions \eqref{Sqq}, \eqref{Sqgq} and \eqref{Sqgqg} in the 
large$-N_c$ limit. Let us first factor out some prefactors:
\be
\f{d\sigma^{q{\cal T}\to qgX}}{d^3kd^3q}=\alpha_S C_F\delta(p^+\!-\!k^+\!-\!q^+) 
M(p,k,q)\ .
\ee
Changing the integration variables to $\textbf{u}\!=\!\textbf{x}\!-\!\textbf{b},$ 
$\textbf{v}\!=\!z\textbf{x}\!+\!(1\!-\!z)\textbf{b},$ 
$\textbf{u}'\!=\!\textbf{x}'\!-\!\textbf{b}',$ and 
$\textbf{v}'\!=\!z\textbf{x}'\!+\!(1\!-\!z)\textbf{b}',$ one writes
\bea
M(p,k,q)=\int\f{d^2u}{(2\pi)^2}\f{d^2u'}{(2\pi)^2}\ e^{i\kappa\cdot(\textbf{u}'-\textbf{u})}
\sum_{\lambda\alpha\beta}
\phi^{\lambda^*}_{\alpha\beta}(p,k^+,\textbf{u}')\phi^{\lambda}_{\alpha\beta}(p,k^+,\textbf{u})
\int\f{d^2v}{(2\pi)^2}\f{d^2v'}{(2\pi)^2}\ e^{i\Delta\cdot(\textbf{v}'-\textbf{v})}\nn\\
\left(e^{-\f{N_c}4[\Gamma(\textbf{u},x_A)+\Gamma(\textbf{u}',x_A)
+\Gamma(\textbf{v}-\textbf{v}'+(1-z)(\textbf{u}-\textbf{u}'),x_A)]}
-e^{-\f{N_c}4[\Gamma(\textbf{u},x_A)+\Gamma(\textbf{v}-\textbf{v}'+(1-z)\textbf{u},x_A)]}
\right.\nonumber\\\left.
-e^{-\f{N_c}4[\Gamma(\textbf{u}',x_A)+\Gamma(\textbf{v}'-\textbf{v}+(1-z)\textbf{u}',x_A)]}
+e^{-\f{N_c}4\Gamma(\textbf{v}-\textbf{v}',x_A)}\right)\label{inter1}
\eea
where we have introduced the following transerve momenta
\be
\kappa=(1-z)k_\perp+z(p_\perp-q_\perp)\ ,\hspace{1cm}\Delta=k_\perp+q_\perp-p_\perp\ .
\ee
In \eqref{inter1}, the transverse momentum transfered during the collision $\Delta$ is Fourier conjugate to $\textbf{v}\!-\!\textbf{v}'$, while $\kappa,$ which caracterizes the invariant mass of the final-state $qg$ system, is Fourier conjugate to $\textbf{u}\!-\!\textbf{u}'.$

We now change the integration variables $\textbf{v}$ and $\textbf{v}'$ to
$\textbf{r}\!=\!\textbf{v}\!-\!\textbf{v}',$ and $\textbf{B}=(\textbf{v}\!+\!\textbf{v}')/2.$
The integration over $\textbf{B}$ represents the impact parameter integration. Following our approximation that in the correlators the $\textbf{B}$ dependence factorizes (and is not explicitely indicated), one has $\int d^2B(1-S)\!=\!S_T\ (1-S)$ and the $\textbf{B}$ integration simply yields the normalization factor $S_T,$ which caracterizes the transverse area of the target. One obtains
\bea
M(p,k,q)=\f{S_T}{4\pi^2}\int\f{d^2u}{(2\pi)^2}\f{d^2u'}{(2\pi)^2}\ 
\sum_{\lambda\alpha\beta}
\phi^{\lambda^*}_{\alpha\beta}(p,k^+,\textbf{u}')\phi^{\lambda}_{\alpha\beta}(p,k^+,\textbf{u})
\int\f{d^2r}{(2\pi)^2}\ e^{-i\Delta\cdot\textbf{r}}e^{-\f{N_c}4\Gamma(\textbf{r},x_A)}\nn\\
\left(e^{i(p_\perp-q_\perp)\cdot(\textbf{u}'-\textbf{u})}e^{-\f{N_c}4[\Gamma(\textbf{u},x_A)
+\Gamma(\textbf{u}',x_A)]}
-e^{i\kappa\cdot \textbf{u}'-i(p_\perp-q_\perp)\cdot \textbf{u}}
e^{-\f{N_c}4[\Gamma(\textbf{u},x_A)]}
\right.\nn\\\left.
-e^{-i\kappa\cdot\textbf{u}+i(p_\perp-q_\perp)\cdot\textbf{u}'}
e^{-\f{N_c}4[\Gamma(\textbf{u}',x_A)]}
+e^{i\kappa\cdot(\textbf{u}'-\textbf{u})}\right)\ .
\eea
Finally, performing the Fourier transforms back to momentum space, the cross-section can be written in the following compact form:
\be
\f{d\sigma^{q{\cal T}\to qgX}}{d^3kd^3q}=S_T\f{\alpha_S C_F}{4\pi^2}
\delta(p^+\!-\!k^+\!-\!q^+) 
\sum_{\lambda\alpha\beta}
\left|I^{\lambda}_{\alpha\beta}(p,k^+,p_\perp\!-\!q_\perp;{x_A})-
\psi^{\lambda}_{\alpha\beta}(p,k^+,\kappa)\right|^2 F_{x_A}(\Delta)
\label{dijet}\ee
where we recall that $\psi^{\lambda}_{\alpha\beta}$ is the $q\!\to\!qg$ wavefunction
\eqref{qtoqgmom} in momentum space, and with 
\be
I^{\lambda}_{\alpha\beta}(p,k^+,\kappa;{x_A})
=\int\f{d^2u}{(2\pi)^2}e^{-i\kappa\cdot\textbf{u}}e^{-\f{N_c}4\Gamma(\textbf{u},x_A)}
\phi^{\lambda}_{\alpha\beta}(p,k^+,\textbf{u})\ .
\ee
We have also introduced the so-called unintegrated gluon distribution $F_{x_A}(\Delta),$ which is simply the Fourier transform of the 2-point function $S_{q\bar q}^{(2)}:$
\be
F_{x_A}(\Delta)=\int\f{d^2r}{(2\pi)^2}\ e^{-i\Delta\cdot\textbf{r}}
e^{-\f{N_c}4\Gamma(\textbf{r},x_A)}\ .
\label{funcF}\ee

It is important to stress that $F_{x_A}(\Delta)$ is not the usual unintegrated distribution defined in the leading-twist approximation of perturbative QCD, but it is rather an all-twist gluon distribution, as multiple scattering are taken into account. Although they coincide for large momentum, $F_{x_A}(\Delta)$ contains more information, as it is also properly defined in the infrared (meaning for $\Delta^2<Q_s^2$). Note that the numerical Fourier transformation in \eqref{funcF} suffers from a positivity problem and features oscillations for large momenta \cite{lm}. This is due to the infrared behavior of the initial condition 
$\Gamma(\textbf{r},x_0),$ and the problem is carried to lower values of $x_A$ by the BK evolution. To cure this, we substituted the $\Theta(k-|\textbf{r}|\Lambda_{QCD})$ function in \eqref{mvmod} by the smooth cutoff function $2\arctan[(k/|\textbf{r}|\Lambda_{QCD})^2]/\pi.$ This modification of the infrared regularisation does not influence the results provided 
$F_{x_A}$ is evaluated for values of $|\Delta|$ much bigger than $\Lambda_{QCD},$ and this will be the case

Finally let us insist that, as already mentioned, the unintegrated gluon distribution 
$F_{x_A}$ is not enough to characterize the CGC, and for the process considered here, more information is needed to compute the cross-section. It is contained in the modified wavefunction
$I^{\lambda}_{\alpha\beta}:$
\be
I^\lambda_{\alpha\beta}(p,k^+,\kappa;{x_A})=
\f1{\sqrt{k^+}}\left\{\begin{array}{lll}\D
\sqrt{2}\varepsilon_\perp^1\cdot G_{x_A}(\kappa\!-\!z p_\perp)
[\delta_{\alpha-}\delta_{\beta-}\!+\!(1\!-\!z)\delta_{\alpha+}\delta_{\beta+}]
\!+\!H_{x_A}(\kappa\!-z\! p_\perp)\delta_{\alpha+}\delta_{\beta-}\hspace{0.5cm}\lambda=1
\\\\\D
\sqrt{2}\varepsilon_\perp^2\cdot G_{x_A}(\kappa\!-z\! p_\perp)
[\delta_{\alpha+}\delta_{\beta+}\!+\!(1\!-\!z)\delta_{\alpha-}\delta_{\beta-}]
\!-\!H_{x_A}(\kappa\!-\!z p_\perp)\delta_{\alpha-}\delta_{\beta+}\hspace{0.5cm}\lambda=2
\end{array}\right.
\ee
given in terms of the functions
\bea
G_{x_A}(\kappa)&=&mz\f{\kappa}{|\kappa|}\int du\ e^{-\f{N_c}4\Gamma(u,x_A)}
uK_1(mzu)J_1(|\kappa|u)\ ,\label{funcG}\\
H_{x_A}(\kappa)&=&mz^2\int du\ e^{-\f{N_c}4\Gamma(u,x_A)}uK_0(mzu)J_0(|\kappa|u)\ ,
\label{funcH}\eea
that contain the rest of the $x_A$ dependence. The functions $F_{x_A},$ $G_{x_A}$ (which is a two-dimensional vector) and $H_{x_A}$ are all obtained from the function $\Gamma(u\!=\!|\textbf{u}|,x_A),$ itself obtained by solving the BK equation.

\section{Azimuthal angle correlations in $d\!-\!Au$ collisions}

In this final section we consider the inclusive two-particle spectrum for the process 
$h{\cal T}\!\to\!h_1h_2X$ and, as an application, we study the correlations in azimuthal angle between the measured particles $h_1$ and $h_2,$ which are both detected at forward rapidities. We consider deuteron-gold collisions at RHIC center-of-mass energies 
($\sqrt{s}\!=\!200\ \mbox{GeV/nucleon}$) and investigate the suppresion of the back-to-back peak as a function of the transverse momenta and rapidities of the two particles.

\subsection{The inclusive quark-gluon production cross-section 
$\sigma^{h{\cal T}\to qgX}$}

Let us denote $P^+$ the momentum of the hadron $h.$ To obtain the cross-section
$\sigma^{h{\cal T}\to qgX}$ from the partonic cross-section $\sigma^{q{\cal T}\to qgX},$
one can use the collinear factorization of the quark density inside the hadron
$q(p^+/P^+,\mu^2)$ where $\mu^2$ is the factorization scale:
\be
\f{d\sigma^{h{\cal T}\to qgX}}{d^3kd^3q}=\int dx\ q(x,\mu^2)\
\f{d\sigma^{q{\cal T}\to qgX}}{d^3kd^3q}(p^+\!=\!xP^+,p_\perp\!=\!0)\ .
\ee
Indeed, we are interested in measurements that probe only large values of $p^+/P^+.$ By contrast, partons with small fraction of momentum are probed inside the target, and it cannot be described by a single gluon distribution (as obvious from formula \eqref{qTtoqgX} for instance). We shall later use the factorization scale $\mu^2\!=\!\Delta^2$ which is the transverse momentum transfered during the collision, and it is supposed to be large to justify our perturbative calculations.

As a function of $k_\perp$ and $\Delta,$ the $\sigma^{h{\cal T}\to qgX}$ cross-section reads
\be
\f{d\sigma^{h{\cal T}\to qgX}}{d^2k_\perp d^2q_\perp dy_k dy_q}=
S_T\f{\alpha_S C_F}{4\pi^2}(1\!-\!z)x_h q(x_h,\Delta^2)
k^+\sum_{\lambda\alpha\beta}
\left|I^{\lambda}_{\alpha\beta}(p,k^+,k_\perp\!-\!\Delta;{x_A})-
\psi^{\lambda}_{\alpha\beta}(p,k^+,k_\perp\!-\!z\Delta)\right|^2 F_{x_A}(\Delta)
\label{hTtoqgX}
\ee
where in $I^{\lambda}_{\alpha\beta}$ and $\psi^{\lambda}_{\alpha\beta},$ $p$ should be understood as $(p^+\!=\!q^+\!+\!k^+,p_\perp\!=\!0).$ Also, $z$ now stands for
$z\!=\!k^+/(k^+\!+\!q^+)$ and we have denoted $x_h\!=\!(k^+\!+\!q^+)/P^+,$ the fraction of momentum of the probed quark inside the incoming hadron. Note that similarly, there is conservation of momentum along the $x^-$ direction and one has $x_A\!=\!(k^-\!+\!q^-)/P^-$ where $P^-$ is the momentum of the incoming target. In terms of the rapidity of the quark
$y_q,$ the rapidity of the gluon $y_q$ and the center of mass energy $\sqrt{s},$ one has:
\be
z=\f{|k_\perp|e^{y_k}}{|k_\perp|e^{y_k}+|q_\perp|e^{y_q}}\ ,\hspace{0.5cm}
x_h=\f{|k_\perp|e^{y_k}+|q_\perp|e^{y_q}}{\sqrt{s}}\ ,\hspace{0.5cm}
x_A=\f{|k_\perp|e^{-y_k}+|q_\perp|e^{-y_q}}{\sqrt{s}}\ .
\label{kine}\ee
It is clear that in order to have $x_h\lesssim 1$ and $x_A\ll1,$ one needs forward rapidities, in the hemisphere in which the hadron $h$ fragments, where $y_q$ and $y_k$ are positive.

In this work, we do not take into account the fragmentation of the final-state quark and gluon into hadrons. Since we are interested in azimuthal angle correlations, fragmentation does not play a important role. Also, we do not include in the process $h{\cal T}\!\to\!h_1h_2X,$ the contributions of the gluon-initiated subprocesses $g{\cal T}\to q\bar qX$ and
$g{\cal T}\to ggX.$ Measurements at forward rapidities at RHIC energies involve values of $x_h$ so high that they are only sensitive to the valence quarks. In the case of the LHC, it is likely that one also needs to account for the gluon-initiated subprocesses.

In order to simplify the numerical computations, we shall work with massless quarks. Using $m=0$
yields
\be
G_{x_A}(k_\perp)=\f{k_\perp}{|k_\perp|}\int du\ e^{-\f{N_c}4\Gamma(u,x_A)}J_1(|k_\perp|u)\ ,
\hspace{1cm}H_{x_A}(k_\perp)=0
\label{nomass}\ee
and in \eqref{hTtoqgX}, the summation over the quark and gluon spins and polarizations becomes
\be
k^+\sum_{\lambda\alpha\beta}
\left|I^{\lambda}_{\alpha\beta}(p,k^+,k_\perp\!-\!\Delta;x_A)-
\psi^{\lambda}_{\alpha\beta}(p,k^+,k_\perp\!-\!z\Delta)\right|^2=2\left[1\!+\!(1\!-\!z)^2\right]
\left|G_{x_A}(k_\perp\!-\!\Delta)-\f{k_\perp\!-\!z\Delta}{|k_\perp\!-\!z\Delta|^2}\right|^2
\ .\label{massless}\ee
In Fig.2, we display the functions $F_{x_A}$ and $G_{x_A}$ (in the massless case) obtained from the BK evolution. More precisely, the function $\Delta^2\ F_{x_A}(\Delta)$ is displayed in Fig.2a and as is well-known, it is peaked for $\Delta^2\!\simeq\!Q_s^2,$ and therefore as $x_A$ decreases the peak travels towards higher momenta. The evolution is however quite slow, considering the small value of the saturation exponent $\lambda.$ The function 
$1\!-\!k_\perp\cdot G_{x_A}(k_\perp)$ is represented in Fig.2b, it decreases from 1 to 0 as
$k_\perp^2$ increases, and the front travels to higher momenta as $x_A$ decreases.

\begin{figure}[t]
\begin{center}
\epsfig{file=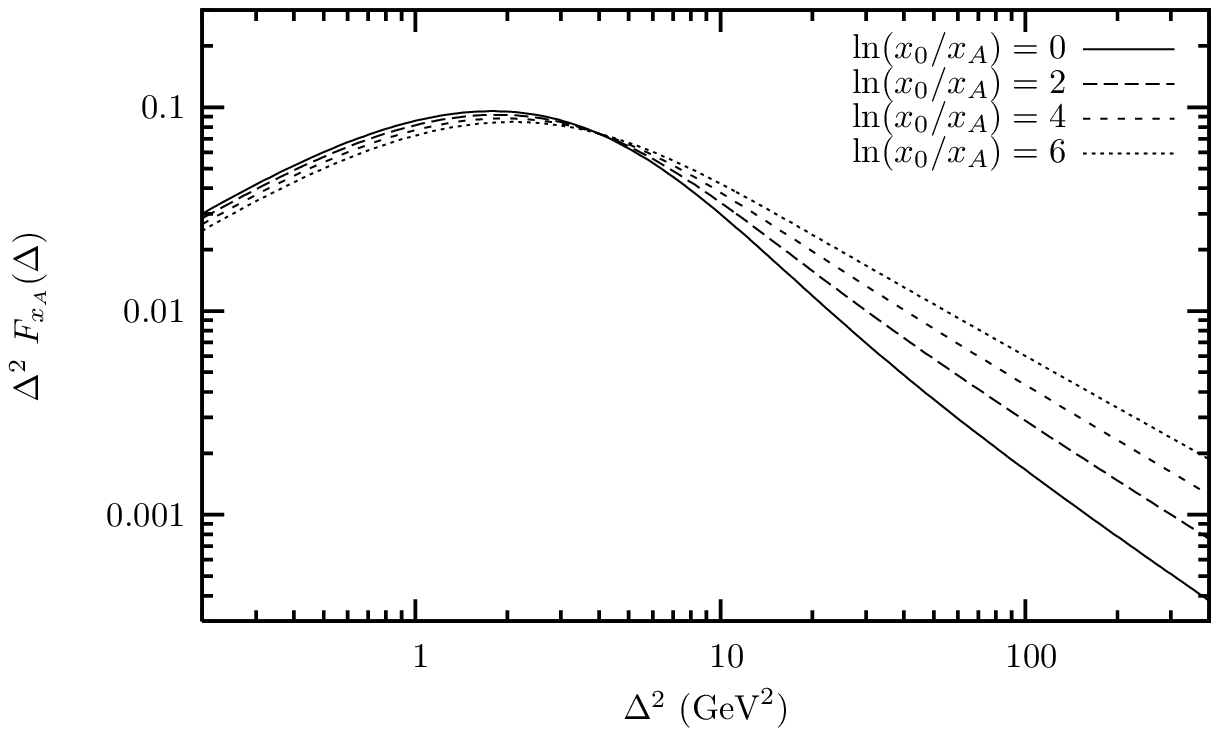,width=8.5cm}
\hfill
\epsfig{file=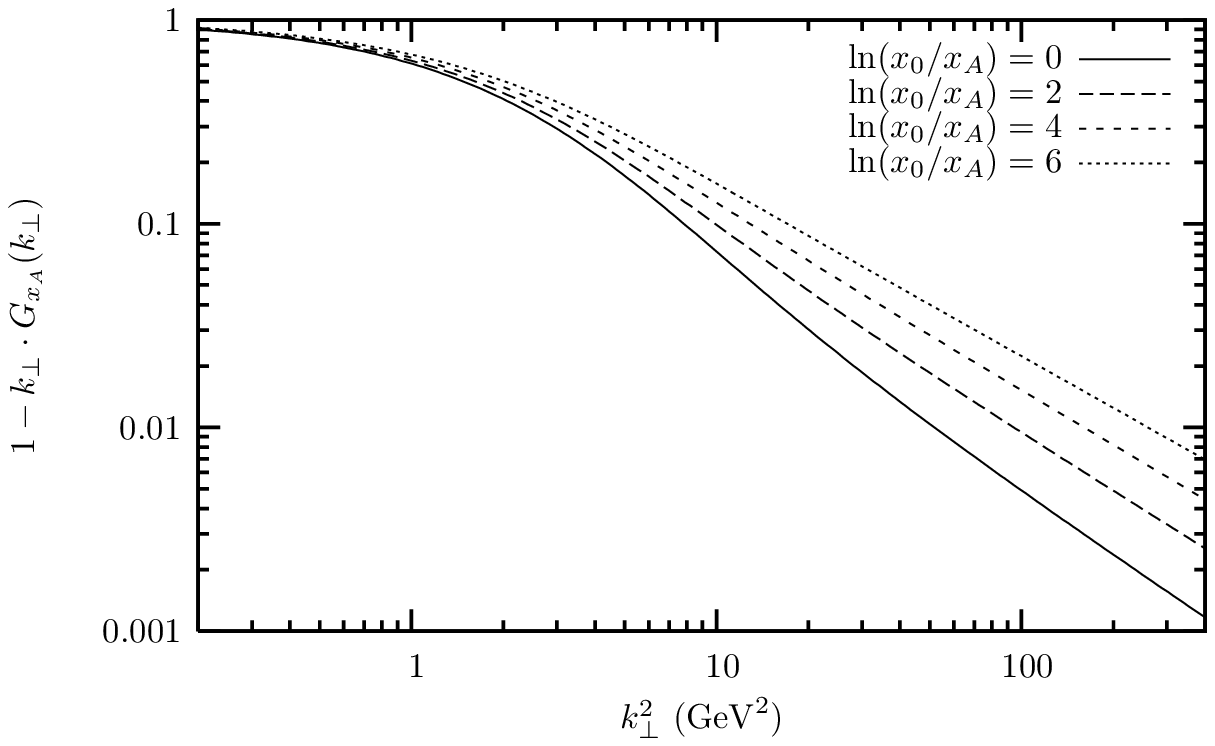,width=8.5cm}
\caption{Left plot: the dimensionless unintegrated gluon distribution 
$\Delta^2\ F_{x_A}(\Delta)$ as a function of $\Delta^2$ (see formula \eqref{funcF}). Right plot: the dimensionless quantity $1\!-\!k_\perp\cdot G_{x_A}(k_\perp)$ as a function of
$k_\perp^2$ (see formula \eqref{nomass}, in the massless case). The curve for
$x_A\!=\!x_0\!=\!0.01$ is obtained from the MV initial condition \eqref{mvmod}. The evolution towards smaller values of $x_A$ is obtained with the BK equation \eqref{bk}, and is shown over 6 units of $\ln(x_0/x_A).$}
\end{center}
\end{figure}

At this point, it is easy to recover the perturbative limit of the inclusive two-particle spectrum \eqref{hTtoqgX}: for $|k_\perp|\!\gg\!Q_s,$ $G_{x_A}(k_\perp)=k_\perp/|k_\perp|^2$ leads to $k_T-$factorization. For $|\Delta|\!\gg\!Q_s$ one has
$F_{x_A}(\Delta)=Q_s^2(x_A)/(\pi\Delta^4)$ \cite{gelpes} which yields
\be
\f{d\sigma^{h{\cal T}\to qgX}_{pQCD}}{d^2k_\perp d^2q_\perp dy_k dy_q}=
S_T\f{\alpha_S C_F}{2\pi^3}\f{(1-z)^3[1+(1-z)^2]Q_s^2(x_A)}
{(k_\perp\!-\!\Delta)^2\Delta^2(k_\perp\!-\!z\Delta)^2}
x_h q(x_h,\Delta^2)\ .
\ee

\subsection{An application: azimuthal angle decorrelations}

We will now use the inclusive two-particle spectrum \eqref{hTtoqgX} to investigate the process
$h{\cal T}\!\to\!h_1h_2X,$ and in particular the cross-section as a function of
$\Delta\phi\!=\!\phi_1\!-\!\phi_2,$ the difference in azimuthal angles of the two measured particles $h_1$ and $h_2.$ We will study the normalized $\Delta\phi$ distribution
\be
\f{1}{\sigma}\f{d\sigma}{d\Delta\phi}\equiv
\lr{\f{d\sigma^{h{\cal T}\to h_1h_2X}}{dp_{T_1}dp_{T_2}dy_1 dy_2}}^{-1}
\f{d\sigma^{h{\cal T}\to h_1h_2X}}{dp_{T_1}dp_{T_2}dy_1 dy_2 d\Delta\phi}
\label{obs}\ee
where $(p_{T_1},\phi_1)$ and $(p_{T_2},\phi_2)$ are the transverse momenta of the measured hadrons and $y_1$ and $y_2$ are their rapidities. Our results can be applied to $d\!-\!Au$ collisions at RHIC, and to compute $x_h$ and $x_A$ we will use
$\sqrt{s}\!=\!200\ \mbox{GeV}.$

A given final-state configuration can be obtained in two possible ways from the 
$h{\cal T}\!\to\!qgX$ process, depending on which particle (1 or 2) comes from the quark and which comes from the gluon. While $x_h$ and $x_A$ are the same in both situations, it is not the case of $z$ (which is changed into $1\!-\!z$), and therefore the cross-section
\eqref{hTtoqgX} is not symmetric with respect to the two situations, as it is a decreasing function of $z.$ Let us choose to label the particles such that
$p_{T_1}e^{y_1}\!>\!p_{T_2}e^{y_2}.$ If the quark (resp. gluon) is the particle 1, then 
$z\!<\!1/2$ (resp. $z\!>\!1/2$). This shows that, for similar transverse momenta 
$p_{T_1}\!\sim\!p_{T_2},$ the favored configuration is the one where the quark is the most forward particle; this is especially true when $y_1\!-\!y_2$ is large. In any case, we take into account both situations.

The massless quark approximation is valid when $(k_\perp\!-\!z\Delta)^2\!\gg\!m^2,$ therefore
we will stay away from the situation $y_1\!=\!y_2,$ because when $\Delta\phi=0$ it leads
$(k_\perp\!-\!z\Delta)^2\!=\!0.$ In this situation the factor $(k_\perp\!-\!z\Delta)^{-2}$ in \eqref{massless} should actually be replaced by $1/m^2;$ there is an increase of the cross-section when $\Delta\phi\!\simeq\!0$ and $y_1\!\simeq\!y_2,$ which corresponds to the quark and the gluon being collinear. Also we shall not consider the situation $p_{T_1}\!=\!p_{T_2}$ which implies that $\Delta^2\!=\!0$ for $\Delta\phi\!=\!\pi.$ Indeed, we would like to work with
$|\Delta|\!\gg\!\Lambda_{QCD}.$

As can be seen from the kinematics \eqref{kine}, the most forward of the two particles essentially determines the value of $x_h$ while the most central one determines the value of $x_A.$ In order to study the effect of the CGC evolution, the ideal situation would be to keep
$x_h$ fixed and to vary $x_A.$ In practice, this is better realized by fixing the rapidity and momentum of the most forward particle and by varying the kinematics of the other. Note that doing the opposite would emphasize the $x_h$ evolution of $q(x_h,\Delta^2),$ rather than focus on the $x_A$ evolution of $F_{x_A}$ and $G_{x_A}.$ Moreover, the cross-section \eqref{hTtoqgX} is quite sensitive to choice of factorization scale in the quark density, so it is better to keep $x_h$ constant. Note that varying the rapidities at fixed $y_1\!-\!y_2$ would keep the product $x_h x_A$ constant, and would force a competition between the evolution of
$q(x_h,\Delta^2)$ with increasing $x_h$ and the CGC evolution with decreasing $x_A.$

In Fig.3a, we have studied the $\Delta\phi$ spectrum \eqref{obs} in the situation in which
$p_{T_1}\!=\!3.5\ \mbox{GeV},$ $p_{T_2}\!=\!2\ \mbox{GeV},$ $y_1\!=\!3.5$ and $y_2$ is varied from $1.5$ to $2.5.$ As $y_2$ increases, the value of $x_A$ decreases and the suppression of the azimuthal correlation is more important. However the effect is quite small, because the increase of the saturation scale with decreasing $x_A$ is rather slow. In Fig.3b, we investigate the situation for which $p_{T_1}\!=\!5\ \mbox{GeV},$ $y_1\!=\!3.5,$ $y_2\!=\!2$ and
$p_{T_2}$ is varied $1.5\ \mbox{GeV}$ to $3\ \mbox{GeV}.$ As $p_{T_2}$ decreases, it gets closer to the saturation scale $Q_s$ (which also slightly increases as $x_A$ decreases), and the suppression of the azimuthal correlation increases. Varying $p_{T_2}$ at fixed $y_2$ allows to probe the ratio $p_{T_2}/Q_s$ over a larger range, so the effect is much bigger than when varying $y_2$ at fixed $p_{T_2}.$

\begin{figure}[t]
\begin{center}
\epsfig{file=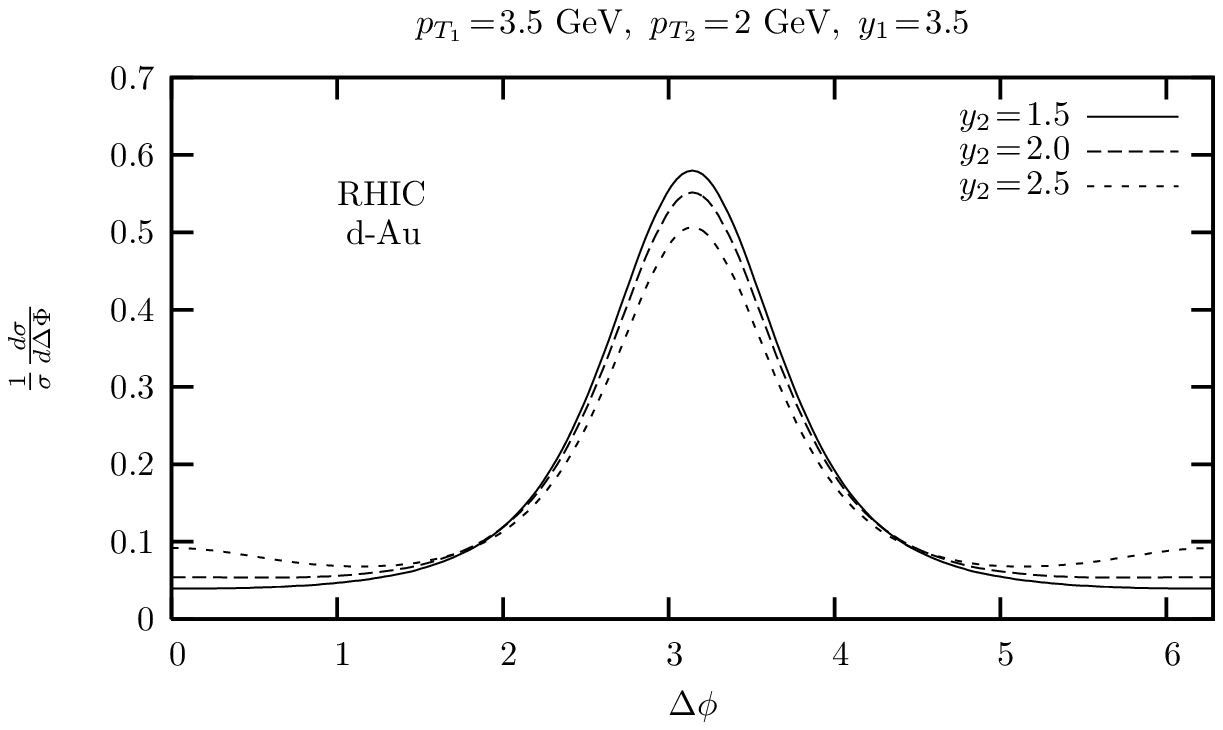,width=8.5cm}
\hfill
\epsfig{file=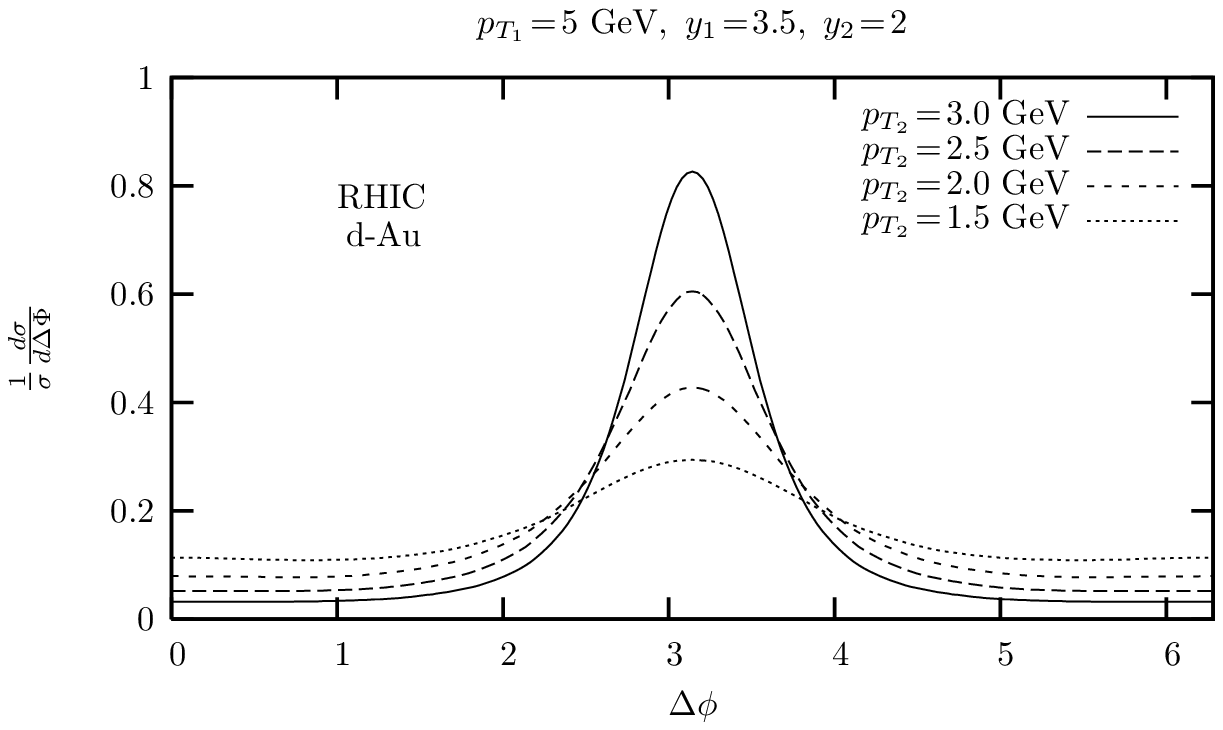,width=8.5cm}
\caption{The $\Delta\phi$ spectrum \eqref{obs} in two situations for the RHIC energy
$\sqrt{s}\!=\!200\ \mbox{GeV/nucleon}.$ Fig.3a: $p_{T_1}\!=\!3.5\ \mbox{GeV},$
$p_{T_2}\!=\!2\ \mbox{GeV},$ $y_1\!=\!3.5$ and $y_2$ is varied from $1.5$ to $2.5.$
Fig.3b: $p_{T_1}\!=\!5\ \mbox{GeV},$ $y_1\!=\!3.5,$ $y_2\!=\!2$ and $p_{T_2}$ is varied
$1.5\ \mbox{GeV}$ to $3\ \mbox{GeV}.$ In both cases, the correlation in azimuthal angle is suppressed as the value of $x_A$ probed in the process decreases. Varying $p_{T_2}$ at fixed $y_2$ is much more efficient as the ratio $p_{T_2}/Q_s$ varies over a larger range.}
\end{center}
\end{figure}

Experimental measurements of two-particle correlations in azimuthal angle have been performed at RHIC in $d\!-\!Au$ collisions \cite{dAudata} by the PHENIX and STAR collaborations. Our predictions for the fully differential cross section are not directly comparable with the data. One would have to carry out a number of integrations over the kinematic variables, while properly taking into account the kinematic cuts applied by the experiments for the different measurements, but this goes beyond the scope of this work. Nevertheless the exploratory measurements of STAR with $\pi^0$ at forward rapidity and charged hadrons at mid rapidity are qualitatively consistent with a suppression of the back-to-back peak with respect to $p\!-\!p$ collisions. By contrast, the measurements of PHENIX do not show any evidence of a suppression of the back-to-back peak, but they probe values of $x_A$ which are bigger than $0.01.$ It may very well be that the CGC picture breaks down for values of $x_A$ bigger than $0.01,$ and it justifies our choice not to start the small$-x_A$ evolution at a higher value. 

In Fig.4, the $\Delta\Phi$ spectrum \eqref{obs} is plotted for the same situations as in Fig.3, but with the LHC heavy-ion energy $\sqrt{s}\!=\!5.5\ \mbox{TeV/nucleon}.$ Assuming similar possibilities for the ALICE detector, compared to RHIC detectors, the final-state kinematics are unchanged, and as a result the values of $x_A$ probed in the process are much smaller at the LHC (typically $x_A\!\sim\!5.10^{-5},$) compared to RHIC (typically $x_A\!\sim\!10^{-3}$).
One sees that in both cases, the azimuthal angle decorrelation behaves as a function of $y_2$ and $p_{T_2}$ as in Fig.3, but as indicated by the vertical scale, the spectrum is globally more suppressed and the peak is also slighlty broader. Let us warn that those conclusions are only qualitative, as our calculation is really only suited for RHIC where the $q\!\to\!qg$ process is predominant. The values of $x_h$ probed at the LHC (typically $x_h\!\sim\!0.02$ compared to $x_h\!\sim\!0.5$ at RHIC) are such that the gluon-initiated processes $g\!\to\!q\bar q$ and $g\!\to\!gg$ (not included in our calculation) will dominate the cross-section.

\begin{figure}[t]
\begin{center}
\epsfig{file=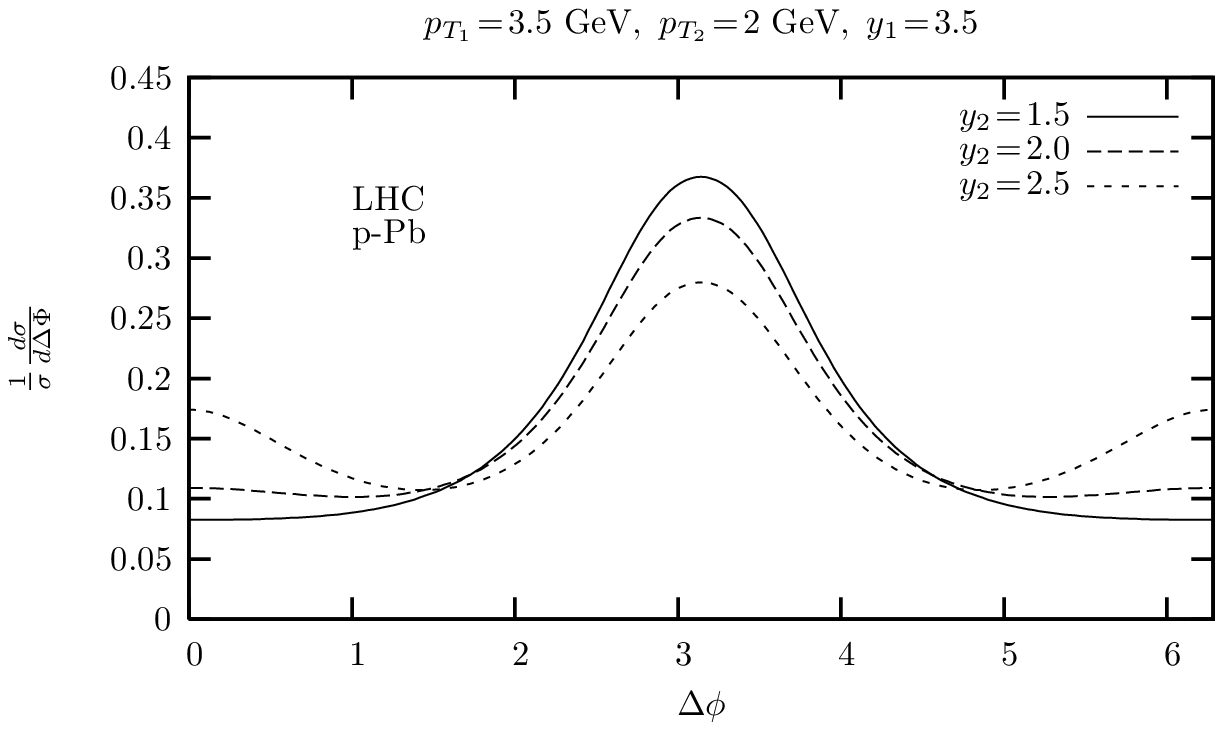,width=8.5cm}
\hfill
\epsfig{file=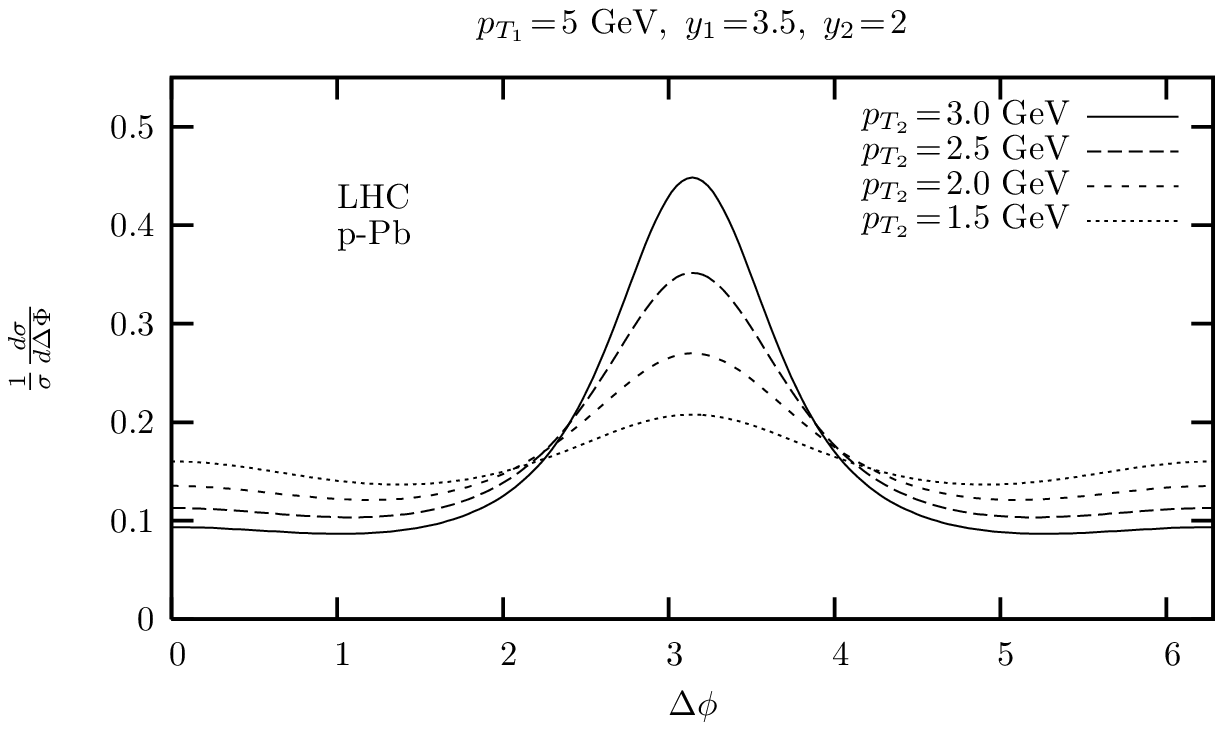,width=8.5cm}
\caption{The $\Delta\phi$ spectrum \eqref{obs} in the two situations studied in Fig.3, but
for the LHC energy $\sqrt{s}\!=\!5.5\ \mbox{TeV/nucleon},$ resulting in probing much smaller values of $x_A.$ In both cases, the correlation in azimuthal angle varies as a function of
$y_2$ and $p_{T_2}$ as in Fig.3, but globally the azimuthal correlation is more suppressed (see the vertical axis) and the peak is less pronounced.}
\end{center}
\end{figure}

\section{Conclusions}

Let us summarize our main results. We computed forward inclusive dijet production
$q{\cal T}\!\to\!qgX$ in the scattering of a quark off a Color Glass Condensate.
The two-particle spectrum \eqref{qTtoqgX} was expressed in terms of correlators
of Wilson lines. With the Gaussian CGC wavefunction \eqref{gauss}, we could compute the correlators in terms of a single function, which in practice is obtained (in the large$-N_c$ limit) by solving the BK equation \eqref{bk} with the MV initial condition \eqref{mvmod}. We applied our expression \eqref{dijet} to the process $h{\cal T}\!\to\!h_1h_2X,$ the inclusive production of two particles $h_1$ and $h_2$ at forward rapidities, in the direction of the dilute hadron $h.$ As an application of formula \eqref{hTtoqgX}, we studied the azimuthal angle correlation in $d\!-\!Au$ collisions. While our results for the fully differential spectrum \eqref{obs} are not yet comparable with the data, we obtain a qualitative agreement.

However, if the CGC discovery at RHIC is to be promoted to the same level than that of the
quark-gluon plasma \cite{rhicexp}, then the qualitative agreements should be made
quantitative. This has been done in the case of single particle production at forward rapidities in $d\!-\!Au$ collisions \cite{dhj}. For azimuthal correlations, this paper
represents a first step, but more efforts are required, on both the experimental and theoretical sides. Other $d\!-\!Au$ runs at RHIC in the future would certainly be of interest,
for instance the accessible range in rapidity could be improved.

The present calculation takes into account the effects of the non-linear QCD evolution at
small$-x.$ However, since the validity of the CGC picture requires $x\!<\!0.01,$ with the
RHIC energy the range probed in $x$ is somewhat limited (down to $10^{-3}$), and the evolution might be only tested at the LHC. In this case the processes $g{\cal T}\!\to\!q\bar qX$ and
$g{\cal T}\!\to\!ggX$ for dijet production should also be included in the calculation. At the level of formula \eqref{qTtoqgX}, the expressions of \cite{nszprl} could be useful, but to proceed further in the derivation, one needs to compute an 8-point function (for the gg final state), which we leave for future work.

Finally, the final-state configuration studied in this paper requires both particles to be produced at forward rapidities, in order to avoid a large rapidity interval between them. The situation considered in \cite{klm} with one particle produced at forward rapidity and the other at mid-rapidity calls for the inclusion of other small$-x$ QCD effects in the BFKL framework \cite{bfkl}. In the context of azimuthal angle correlations, these effects have been the focus of devoted studies \cite{bfklfj,bfklmnj}, however combining them with the CGC evolution included here is still an open and interesting problem.

\begin{acknowledgments}

I am grateful to Francois G\'elis for providing his code that evolves the MV model with the
BK equation. I also thank Raju Venugopalan for reading the manuscript and making valuable comments. This research was supported by RIKEN, Brookhaven National Laboratory and the U.S. Department of Energy [DE-AC02-98CH10886].

\end{acknowledgments}

\end{document}